# Schwarzschild-like metric and a quantum vacuum

P. R. Silva – Retired Associate Professor – Departamento de Física – ICEx – Universidade Federal de Minas Gerais – prsilvafis@gmail.com

ABSTRACT

A quantum vacuum, represented by a viscous fluid, is added to the Einstein's vacuum, surrounding a spherical distribution of mass. This gives as a solution, in spherical coordinates, a Schwarzschild-like metric. The plot of $g_{00}$ and $g_{11}$ components of the metric, as a function of the radial coordinate, display the same qualitative behavior as that of the Schwarzschild metric. However, the temperature of the event horizon is equal to the Hawking temperature multiplied by a factor of two, while the entropy is equal to half of the Bekenstein one.

1- INTRODUCTION

Einstein's theory of the general relativity has as one of its simplest applications, the case of determining the space-time structure in the neighborhood of an isolated, static, uncharged and non-rotating spherical mass M. In these conditions, the solutions of Einstein's equations were developed by Karl Schwarzschild [1] in 1916. A detailed calculation of vacuum spherical symmetric Einstein's equations leading to the Schwarzschild space-time can be found in a book by McMahon [2]. We refer also to another treatment of this problem reported by Schmude [3].

In this work, we first intend to "derive" the Schwarzschild metric by using special relativity, the equivalence principle, and the Newtonian law of gravity. However, the present treatment is somewhat different from others simple derivations of this subject, as are those discussed by Sacks and Ball [4]. Besides this, we propose a modification of the Schwarzschild problem, where the contribution of a viscous' fluid (a kind of quantum vacuum) acting on the test particle used to probe the gravitational field of the source M, will also be considered.

2- MAXIMUM FLOATING PRINCIPLE AND THE SCHWARZSCHILD RADIUS

Derek Paul in studying the dispersion relation of the de Broglie waves [5], has used the equivalence principle in order to analyze the motion of a heavy (guided) photon in the gravitational field. Inspired in Paul's work [5], we developed an alternative way of estimating the Schwarzschild radius of a mass M.

Let us suppose a test particle of mass m probing the gravitational field of a source particle of mass M. The potential energy, U, of the system reads

$$U = - G M m / r. \qquad (1)$$

Next, by using the definition of the de Broglie frequency and the equivalence principle, it is possible to write

$$\hbar \, \delta\omega = \delta U = (G M m / r^2) \, \delta r. \qquad (2)$$

Now let us divide (2) by the zero-point energy of the "de Broglie oscillator related to the test mass". We have

$$\delta U / (½ \, m \, c^2) = 2 \, G M / (c^2 \, r_0) \, (\delta r / r_0). \qquad (3)$$

Upon to identify ½ m $c^2$ with $\hbar\omega_0$, we get from (3) the relation

$$\delta\omega / \omega_0 = [2 \, G M / (c^2 \, r_0)] \, (\delta r / r_0). \qquad (4)$$

Based in (4) it is possible to write

$$\Delta\omega / \omega_0 = [2 \, G M / (c^2 \, r_0)] \, (\Delta r / r_0). \qquad (5)$$

It seems to be the place to invoke the "maximum floating principle", or MFP, namely to seek for a value of $r_0$ such that, simultaneously:

$$| \Delta\omega / \omega_0 | = | \Delta r / r_0 | = 1. \qquad (6)$$

Making the above requirement we find that

$$r_0|_{MFP} = 2 \, G M / c^2 \equiv R_S, \qquad (7)$$

where $R_S$ stands for the Schwarzschild radius.

3 – THE SCHWARZSCHILD METRIC

Taking into account the ideas which lead to equations (2), (3), and (4), we write

$$d\omega / \omega = (G M / c^2) \, r^{-2} \, dr. \qquad (8)$$

Now we perform the integration

$$\int_L^H (d\omega/\omega) = (GM/c^2)\int_R^r (r^{-2}\,dr). \tag{9}$$

In (9), we take $L = \omega_0$, and $H = \omega$, and pursuing further we get

$$\omega = \omega_0 \exp[(-GM/c^2)(1/r - 1/R)], \tag{10}$$

and

$$\omega^2 = \omega_0^2 \exp[(-2GM/c^2)(1/r - 1/R)]. \tag{11}$$

Making the choice: $R = R_S$, we get

$$\omega^2 = \omega_0^2 \exp(1 - R_S/r). \tag{12}$$

Now we construct the auxiliary metric

$$d\sigma^2 = \omega^2\,dt^2 - k^2\,dr^2 - r^2(d\theta^2 + \sin^2\theta\,d\varphi^2). \tag{13}$$

In (13) we take $k^2$, such that

$$k^2/k_0^2 = \omega_0^2/\omega^2. \tag{14}$$

Relation (14) is a reminiscence of the time dilation and space contraction of the special relativity.

But we seek for a metric which becomes flat in the limit of weak fields, namely as $G \to 0$, or $r \to \infty$. This can be accomplished by defining

$$w^2 = \ln(\omega^2/\omega_0^2), \quad \text{and} \quad \kappa^2 = 1/\omega^2. \tag{15}$$

Making the above choices we can write

$$ds^2 = (1 - R_S/r)\,dt^2 - (1 - R_S/r)^{-1}\,dr^2 - r^2(d\theta^2 + \sin^2\theta\,d\varphi^2). \tag{16}$$

Relation (16) is the well known Schwarzschild metric or Schwarzschild space-time.

4 - A QUANTUM VACUUM CONTRIBUTION

Let us suppose that the vacuum surrounding a source mass M behaves as a viscous medium. Therefore the test particle of mass m which probes the gravitational field of the source mass M, experiments a force proportional to the particle velocity v. By taking separately the effect of this force on the test particle, and using Newton's second law of motion, we can write

$$m\, dv/dt = -p/\tau, \qquad (17)$$

where $\tau$ is the characteristic time of the problem. Multiplying (17) by v, and considering that

$$v\, dt = dr, \quad \text{and} \quad p = \hbar/r, \qquad (18)$$

we have

$$dK_{qt} = -(\hbar/\tau)(dr/r). \qquad (19)$$

Integrating (19), we obtain

$$\Delta K_{qt} = -(\hbar/\tau)\ln(r/R). \qquad (20)$$

In (20), $\Delta K_{qt}$ is the decrease in the kinetic energy of the test particle due to the work of the viscous force, and R is some radius of reference.
  Associated to $\Delta K_{qt}$, we define a potential energy given by

$$\Delta U_{qt} = -\Delta K_{qt} = (\hbar/\tau)\ln(r/R). \qquad (21)$$

Let us take the total potential energy, U, as

$$U = U_{gr} + \Delta U_{qt} = -GMm/r + (\hbar/\tau)\ln(r/R). \qquad (22)$$

Now we use the equivalence principle, namely

$$\delta U = (GMm)(dr/r^2) + (\hbar/\tau)(dr/r) = \hbar\,\delta\omega. \qquad (23)$$

In the next step we divide $\delta U$ by $mc^2$, and use the de Broglie relation $\hbar\omega = mc^2$, getting:

$$\delta\omega/\omega = (GM/c^2)(dr/r^2) + [\hbar/(mc^2\tau)](dr/r). \qquad (24)$$

Upon integrating between $\omega_0$ and $\omega$, and from R to r, yields

$$\ln(\omega/\omega_0) = -GM/(rc^2) + [\hbar/(mc^2\tau)]\ln(r/R). \qquad (25)$$

Next, we make the choice of a very special test mass satisfying the relation

$$G M m = G M_P^2 = \hbar c, \qquad (26)$$

where $M_P$ is the Planck's mass.

Relation (26) implies this test mass has a Compton length equals to half of the Schwarzschild radius of the source mass M. Maybe relation (26) keeps a connection with physics theories which includes a fifth dimension, as discussed by Wesson [6].

Using (26) into (25), we have

$$\ln(\omega/\omega_0) = - G M/(r c^2) + G M/(c^2 c\tau) \ln(r/R), \qquad (27)$$

and

$$\omega^2 = \omega_0^2 \exp\{- R_S/r + [R_S/(c\tau)] \ln(r/R)\}. \qquad (28)$$

In (28), $c\tau$ is the mean free path of the of the test particle in the quantum vacuum (here considered as a viscous medium).
Defining

$$c\tau = R_S, \quad \text{and} \quad r/R = e\, r/R_S, \qquad (29)$$

we get

$$\omega^2 = \omega_0^2 \exp[- R_S/r + \ln(e\, r/R_S)]. \qquad (30)$$

Now, the change of coordinates used in section 3, which led accurately to the Schwarzschild metric, gives some confidence in applying the same transformation to the present case. Therefore we define again:

$$w^2 = \ln(\omega^2/\omega_0^2), \quad \text{and} \quad \kappa^2 = 1/\omega^2, \qquad (31)$$

which leads to:

$$ds^2 = [\ln(e\, r/R_S) - R_S/r] dt^2 - [\ln(e\, r/R_S) - R_S/r]^{-1} dr^2 - r^2 d\Omega^2, \qquad (32)$$

where $d\Omega$ is the differential of a solid angle.
Therefore, by considering this kind of quantum vacuum (a viscous medium), the components of the metric tensor assumes a zero value in the $g_{tt}$ case and acquires a singularity in the $g_{rr}$ case, just at $r = R_S$ (the Schwarzschild radius). We observe that, in order to recover the flat metric in the asymptotic limit, we must take both limits $G \to 0$ and $\hbar \to 0$.

Remember we need to go back to formula (25), as a means to see the quantum contribution to the coefficient of the logarithmic term.

In figure 1, we plot $g_{tt} \equiv g_{00}$ and $g_{rr} \equiv g_{11}$, as a function of $x = r/R_S$, both for the Schwarzschild and Schwarzschild-like metrics. As we can see from figure 1, the two metrics display the same qualitative behavior.

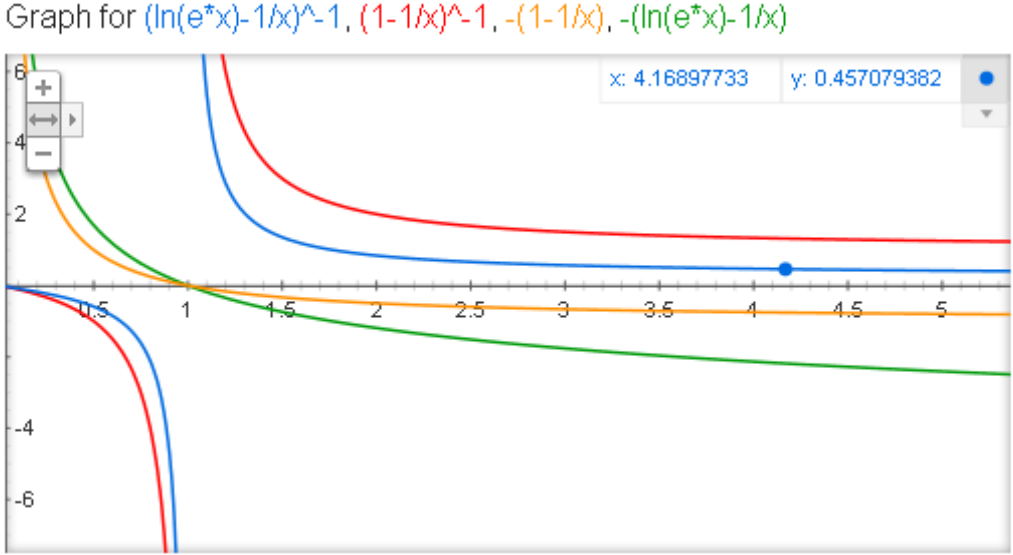

Figure 1 – Plot of components of Schwarzschild and Schwarzschild-like metrics, where $x = r/R_S$.

## 5 – HAWKING-LIKE TEMPERATURE AND BEKENSTEIN-LIKE ENTROPY

Next we derive the Hawking-like temperature [7] and the Bekenstein-like [8] entropy. We start from the solution for the gravitational field with the addition of a quantum vacuum (here represented by a viscous medium). We follow a procedure discussed by Zee [9], please see also [10].

By setting $t \rightarrow -i\tau$, we perform Wick rotation on the metric given by eq. (32), obtaining

$$ds^2 = -\{[\ln(e\, r/R_S) - R_S/r]\, d\tau^2 + [\ln(e\, r/R_S) - R_S/r]^{-1} dr^2 + r^2\, d\Omega^2\}. \quad (33)$$

Now we make the change of coordinates

$$R\, d\alpha = [\ln(e\, r/R_S) - R_S/r]^{1/2}\, d\tau, \quad (34)$$

and

$$dR = [\ln(e\, r/R_S) - R_S/r]^{-1/2}\, dr. \quad (35)$$

We can write (35) as

$$dR = r^{1/2} [ r \ln(e\, r/R_S) - R_S ]^{-1/2} dr, \qquad (35A)$$

or

$$dR \approx R_S^{1/2} [ r \ln(e\, r/R_S) - R_S ]^{-1/2} dr. \qquad (36)$$

Defining

$$r \ln(e\, r/R_S) - R_S = u \qquad (37)$$

implies in

$$du = [ 1 + \ln(e\, r/R_S) ] dr \approx 2 dr. \qquad (38)$$

Inserting (38) into (36) and integrating, we have

$$R \approx R_S^{1/2} u^{1/2}. \qquad (39)$$

Next we write

$$R\, d\alpha \approx R_S^{-1/2} u^{1/2} d\tau, \qquad (40)$$

and upon integrating, taking the limits of $\alpha$ between 0 and $2\pi$ and of $\tau$ between 0 and $\beta$, we get

$$R\, 2\pi \approx R_S^{-1/2} u^{1/2} \beta. \qquad (41)$$

By using (39) and (41), we finally obtain

$$1/\beta \equiv T = 1/(2\pi R_S) = 1/(4\pi G M). \qquad (42)$$

Therefore the temperature, T, of the event horizon of this model is given by twice the Hawking temperature [7].

As a means to obtain the Bekenstein-like entropy let us take the free energy of a black hole as the thermodynamic relation

$$F = E - TS, \qquad (43)$$

where E and F are respectively the internal and free energies and S is the entropy. Let us take the extremum of F, by considering an isothermal process. We have

$$\Delta F = \Delta E - T\Delta S = 0. \tag{44}$$

To evaluate $\Delta E$, we consider a black hole interacting with itself and think about a body of reduced mass equal to M/2. With these ideas in mind we can write

$$\Delta E = \tfrac{1}{2} Mc^2. \tag{45}$$

Substituting T, given by (42) and $\Delta E$, given by (45), in (44) and after solving for $\Delta S$, we obtain

$$\Delta S = (2\pi GM^2), \tag{46}$$

or

$$\Delta S = (1/8)(4\pi R_S^2)/L_P^2, \tag{47}$$

Where $L_P^2 = G$, is the square of the Planck radius. Taking

$$S = S_0 + \Delta S, \tag{48}$$

and by setting $S_0 = 0$, we get the entropy equal to half of the Bekenstein-Hawking one.

## 6 – CONCLUSIONS

In a recent paper S. T. Hong studied the five-dimensional Schwarzschild black hole [11], looking at their geometrical and hydrodynamic aspects. One of the results obtained by him is the Hawking temperature of this five-dimensional manifold, which differs from the four-dimensional one. Making things more explicit, let us write

$$T^H_{d=5} = \{2\pi\, r_H\, [1 - (r_H/r)^2]^{1/2}\}^{-1}, \tag{49}$$

and

$$T^H_{d=4} = \{4\pi\, r_S\, [1 - (r_S/r)]^{1/2}\}^{-1}. \tag{50}$$

We notice that result (49) of Hong [11], approaches the result of the present work (please see eq. (42)), if we take the limit $r \to \infty$, and $r_H = R_S$.

Finally it would be worth to comment that a five-dimensional Schwarzschild-like solution of Einsteins' vacuum was discussed by Steven

Millward in a 2006 paper [12]. As was pointed out by him, certain aspects of his solution for 5-d were found very different from the 4-d Schwarzschild solution, but the geodesics of the two solutions being essentially the same in regions accessible to observation.

ACKNOWLEDGEMENT

I am grateful to my friend Dr. Nilton Penha Silva, for reading a previous version of this manuscript, making interesting suggestions, which help me to improve this work.